\documentclass[twocolumn,superscriptaddress]{revtex4-1}

\usepackage{graphicx}
\usepackage{amsmath}
\usepackage{amssymb}
\usepackage{cleveref}
\usepackage{natbib}
\usepackage[dvipsnames]{xcolor}

\begin{document}

\title{Measuring solar neutrinos over Gigayear timescales with Paleo Detectors}
\author{Natalia Tapia-Arellano}
\email{ntapiaa@vt.edu}
\affiliation{Center for Neutrino Physics, Department of Physics, Virginia Tech, Blacksburg, Virginia 24061, USA }
\author{Shunsaku Horiuchi}
\email{horiuchi@vt.edu}
\affiliation{Center for Neutrino Physics, Department of Physics, Virginia Tech, Blacksburg, Virginia 24061, USA }

\date{\today}

\begin{abstract}
Measuring the solar neutrino flux over gigayear timescales could provide a new window to inform the Solar Standard Model as well as studies of the Earth's long-term climate. We demonstrate the feasibility of measuring the time-evolution of the $^8$B solar neutrino flux over gigayear timescales using paleo detectors, naturally occurring minerals which record neutrino-induced recoil tracks over geological times. We explore suitable minerals and identify track lengths of 15--30 nm to be a practical window to detect the $^8$B solar neutrino flux. A collection of ultra-radiopure minerals of different ages, each some 0.1 kg by mass, can be used to probe the rise of the $^8$B solar neutrino flux over the recent gigayear of the Sun's evolution. We also show that models of the solar abundance problem can be distinguished based on the time-integrated tracks induced by the $^8$B solar neutrino flux. 
\end{abstract}

\maketitle

\section{Introduction}

The concept of ``Solar Activity" is used by the scientific community 
to study the fact that the Sun is not a static quiet object and, in contrast to what was thought in ancient times, evolves \cite{Usoskin:2008bj}. Studying and understanding solar activity not only benefits science itself, but affects directly our environment as living beings and has broad implications: strong solar activity affects directly our communications and travel systems and of course, our planet's climate \cite{2001Sci...294.2130B,2000QSRv...19..403B,Ribas_2005}.

Fundamental to the study of the evolution of the Sun is the Solar Standard Model (SSM), a theoretical tool to investigate the solar interior. First attempts to build a solar model to understand and predict more accurately its core central temperature and to estimate the rates of solar neutrinos started in the mid 20th century, and culminated in the SSM by Bahcall et al.~\cite{PhysRevLett.20.1209}. 
Over the decades, the SSM has been developed and stringently tested with a variety of measurements, not only neutrinos but also helioseismology (for recent reviews, see, e.g., \cite{doi:10.1146/annurev-astro-081811-125539}). 

Initial measurements of the solar neutrinos resulted in less neutrinos than predicted by the SSM, known as the ``solar neutrino problem.'' Many solutions were developed based on producing a cooler solar core, but the phenomenon of neutrino oscillations \cite{Pontecorvo:1967fh}
proved the correct solution with the discovery of the matter-induced Mikheyev-Smirnov-Wolfenstein mechanism  \cite{Mikheev:1986gs, PhysRevD.20.2634, PhysRevD.17.2369, Mikheev:1986wj} and direct measurement of all neutrino flavours of solar neutrinos by the Sudbury Neutrino Observatory (SNO) experiment \cite{PhysRevLett.89.011301, RevModPhys.88.030502}. 
Presently, much of the solar neutrino flux has been experimentally verified (see, e.g., \cite{doi:10.1146/annurev-astro-081811-125539,Wurm:2017cmm}). 
In addition, helioseismic observations probe important properties of the solar interior, such as the sound speed profile, depth of the convective envelope, and surface helium abundance, to percent or better precision (e.g., \cite{DeglInnocenti:1996uex,Gough:1996am}). As a result, the solar structure is now very well constrained. By now, solar neutrinos and helioseismology provide a clear view of the present energy generation rate of the Sun. By contrast, photons reveal the power generation roughly $10^4$--$10^5$ years ago, corresponding to the energy transport timescale within the solar interior \cite{1992ApJ...401..759M}. Within measurement uncertainties, the neutrino and photon observations match, revealing the stability of the solar activity on $\sim 10^5$ year scales.

However, recent measurements of the solar elemental abundances (metallicity) by Asplund et al.~2009 (hereafter AGSS) \cite{Asplund:2009fu} have caused a new conflict within the SSM. The new photospheric measurements indicate that the solar metallicity is lower than previously estimated by Grevesse \& Sauval 1998 (hereafter GS) \cite{Grevesse:1998bj}. The SSM is sensitive to transitions in metals (used to refer to anything above helium), which are an important contributor to opacity. Lower metallicity is associated with a cooler solar core, and in this way, affects the solar interior. Solar models using the new metallicity are no longer able to reproduce helioseismic results, causing the so-called ``solar abundance problem'' \cite{doi:10.1146/annurev-astro-081811-125539, Serenelli_2009}. Extended SSMs, where some of the assumptions of the SSM are relaxed, have been explored as potential solutions, e.g., early accretion causing chemical inhomogeneities \cite{Serenelli_2011}. While the debate continues, a pragmatic approach has been to use two sets of SSMs with different metallicities.  
Neutrinos are one way to observationally test the solar abundance problem through their dependence on the metallicity of the stellar interior. A lower metallicity causes a lower core temperature which in turn has a strong impact on neutrino fluxes, in particular those involving nuclei in their production chain, e.g., $^8$B, CNO neutrinos.

Detecting weakly-interacting particles like solar neutrinos requires immense volumes to instrument enough target numbers to have an appreciable number of neutrino interactions. The same difficulties faced by neutrino measurements are encountered in direct searches for dark matter, where experiments seek to observe the scattering of target nuclei by an incoming dark matter particle. 
Recently, it has been proposed that rock crystals deep in the Earth could act as a new method to detect dark matter \cite{ PhysRevD.99.043014,PhysRevD.99.043541,Edwards:2018hcf,BAUM2020135325}. Called ``paleo detectors'', the idea is that dark matter interactions will leave permanent  
structure or chemical changes (tracks) caused by the recoiling nuclei within the rocks. By strategically collecting samples from geologically quiet areas, the tracks can survive geological periods and measured in the lab through small angle X-ray scattering. And by sampling from deep underground, backgrounds caused by cosmic-ray interactions in the Earth atmosphere can be mitigated. An unavoidable background comes from radioactivity of the Earth, but this can be rejected based on their rich track signatures. 
Furthermore, paleo detectors have competitive exposures compared to terrestrial experiments. Exposure is the product of the target mass and duration in time. 
In the case of paleo detectors, $\sim 1$ Gyr old samples are possible, so even a small mass of $10\,\text{mg}$ would be competitive with a terrestrial experiment of $10^3\,\text{kg}$ running for 10 years: both have exposures $\varepsilon = 0.01\,\text{kg Myr}$. 

In this paper, we consider the detectability of solar neutrinos using paleo detectors. The energies of solar neutrinos are about $\sim 1\, \text{MeV}$, which translate to $\sim \text{keV}$ of recoil energy. This is comparable to the recoil energies $E_R = 0.1 - 100 \, \text{keV}$ caused by dark matter, meaning solar neutrinos should also give rise to damage tracks. Importantly, paleo detectors not only open a new way to search for solar neutrinos, they allow us to probe the Sun in the past on Gyr time scales, something that is not possible with terrestrial detectors. Recently, paleo detectors have been considered also as detectors of supernova neutrinos \cite{PhysRevD.101.103017} and atmospheric neutrinos \cite{Jordan:2020gxx} over similar geological times. 
To investigate the study of the SSM with paleo detectors, we compute two SSMs that differ in their metallicity. We focus on the $^8$B neutrino flux, which shows strong dependence on the solar interior temperature and metallicity models. The $^8$B neutrinos are also among the highest in energy, making them easier to detect compared to other higher-flux but lower-energy solar neutrino components.

This paper is organized as follows. In Section II, we go through our computation of the history of the solar neutrino flux 
with different metallicities which give us the time dependency of the flux. Section III goes through the track formation process, mineral choice, and a basic mathematical framework to compute the estimated number of tracks for each SSM. In Section IV, we use the time-dependent track estimates to check the behaviour or variation in time for our two metallicity models. Finally in Section V we discuss our results and conclude.

\section{Solar Neutrino \& Metallicity Models}\label{sec:solar}

Solar neutrinos arise from multiple reactions within the solar core. We focus on the $^8$B neutrinos due to their higher neutrino energies which facilitates detection, and also for their strong dependence on the solar core temperature, approximately $\propto T^{24}$ (by contrast, the dominant $pp$ chain has a $\propto T^4$ dependence) \cite{Bahcall:1996vj}. Since most of the $^8$B flux is emitted from the interior 5--10\% of the solar radius, the $^8$B neutrinos provide a probe of the solar interior. It is sensitive to the interior metal abundance, due to the metallicity's influence on the electromagnetic opacity and hence temperature gradient inside the Sun. 

We model the Sun using the {\tt MESA} code version r12115 \cite{Paxton:2010ji,Paxton:2013pj,Paxton:2015jva,Paxton:2017eie,Paxton:2019aaa}, closely following the procedure outlined in Farag et al.~2020 \cite{Farag:2020nll}. We adopt their solar models calibrated to reproduce the present day neutrino fluxes, with final age $t_\odot = 4.568$ Gyr \cite{2010NatGe...3..637B}, radius $R_\odot = 6.9566 \times 10^{10}$ cm \cite{2016AJ....152...41P}, photon luminosity $L_\gamma = 3.828 \times 10^{33}$ erg/s \cite{2016AJ....152...41P}, and surface metallicity $Z/X$. These are obtained by using the built-in {\tt MESA} simplex module to vary iteratively the mixing-length parameter and the initial hydrogen, helium, and metal compositions, including the effects of element diffusion \cite{Paxton:2017eie}. The helioseimic signals of these models have been confirmed to be similar to others in the literature \cite{Villante:2013mba}. We refer the reader to Ref.~\cite{Farag:2020nll} for further details of the calculations.

\begin{figure}[t]
    \centering
    \includegraphics[angle=0,width=.48\textwidth]{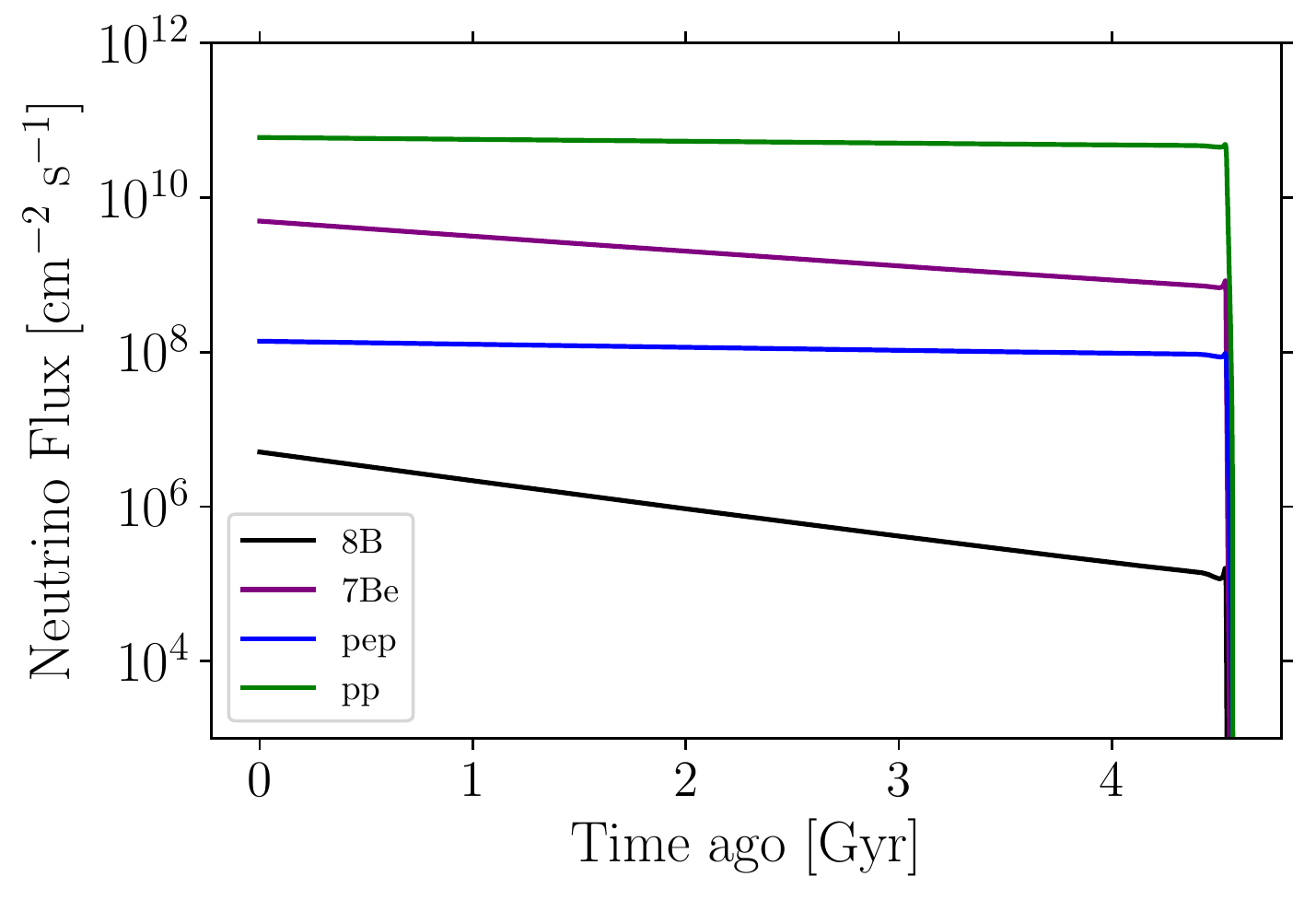}
    \caption{Neutrino flux variation over time, for different components of the solar neutrino flux, for the GS metallicity model. 
    } 
    \label{fig:neutrino_flux}
\end{figure}

We compute two SSMs made to match two different abundance of heavy metals at the surface of the Sun: $Z/X=0.0229$ for GS \cite{Grevesse:1998bj} and $Z/X=0.0181$ for AGSS \cite{Asplund:2009fu}. We use the built-in {\tt MESA} reaction network {\tt add\_pp\_extras} to model the reactions emitting neutrinos, including $pp$, $pep$, $^7$Be, and $^8$B. For the CNO chain, we use the {\tt add\_cno\_extras} and {\tt add\_hot\_cno} and compute the neutrinos from $^{13}$N, $^{15}$O, and $^{17}$F. Our obtained neutrino fluxes for the two metallicity models, compared to experimentally measured values, are listed in Table \ref{tab:flux}. Our predicted fluxes are similar to those in the literature \cite{Robertson:2012ib,Villante:2013mba,Farag:2020nll}. In Figure \ref{fig:neutrino_flux}, we show the time evolution of these neutrino fluxes, as a function of years in the past. We show results only for the GS metallicity model, as the AGSS  is very similar on these scales.

\begin{table}[h]
\begin{tabular}{llll}
\hline
Channel     & GS   & AGSS      & Measurement \\
\hline
$\Phi_{\rm pp}$     & 5.98      & 6.01      & $6.05 (1^{+0.003}_{-0.011})$ \\
$\Phi_{^7{\rm Be}}$    & 4.95      & 4.71      & $4.82 (1^{+0.05}_{-0.04})$ \\
$\Phi_{^8{\rm B}}$     & 5.09      & 4.62      & $5.00(1 \pm 0.03)$ \\
$\Phi_{\rm CNO}$      & 5.12      & 3.92      & $7.0 (1^{+0.43}_{-0.29})$ \\
\hline 
\end{tabular}
\caption{Observed solar neutrino fluxes at Earth, predicted by {\tt MESA} and measured, in units of [cm$^{-2}$ s$^{-1}$], with the following scales: $10^{10}$ ($pp$), $10^9$ (7Be), $10^6$ (B), and $10^8$ (CNO). Measurements are from Ref.~\cite{Bellini:2011rx} as summarized in Refs.~\cite{Robertson:2012ib,Villante:2013mba,Farag:2020nll}, and updated by Borexino's measurement of the CNO flux \cite{Agostini:2020mfq}.}
\label{tab:flux}
\end{table}

\section{Rates per track length spectrum}\label{sec:formalism}

\begin{figure*}[t]
    \centering
    \includegraphics[angle=0,width=.49\textwidth]{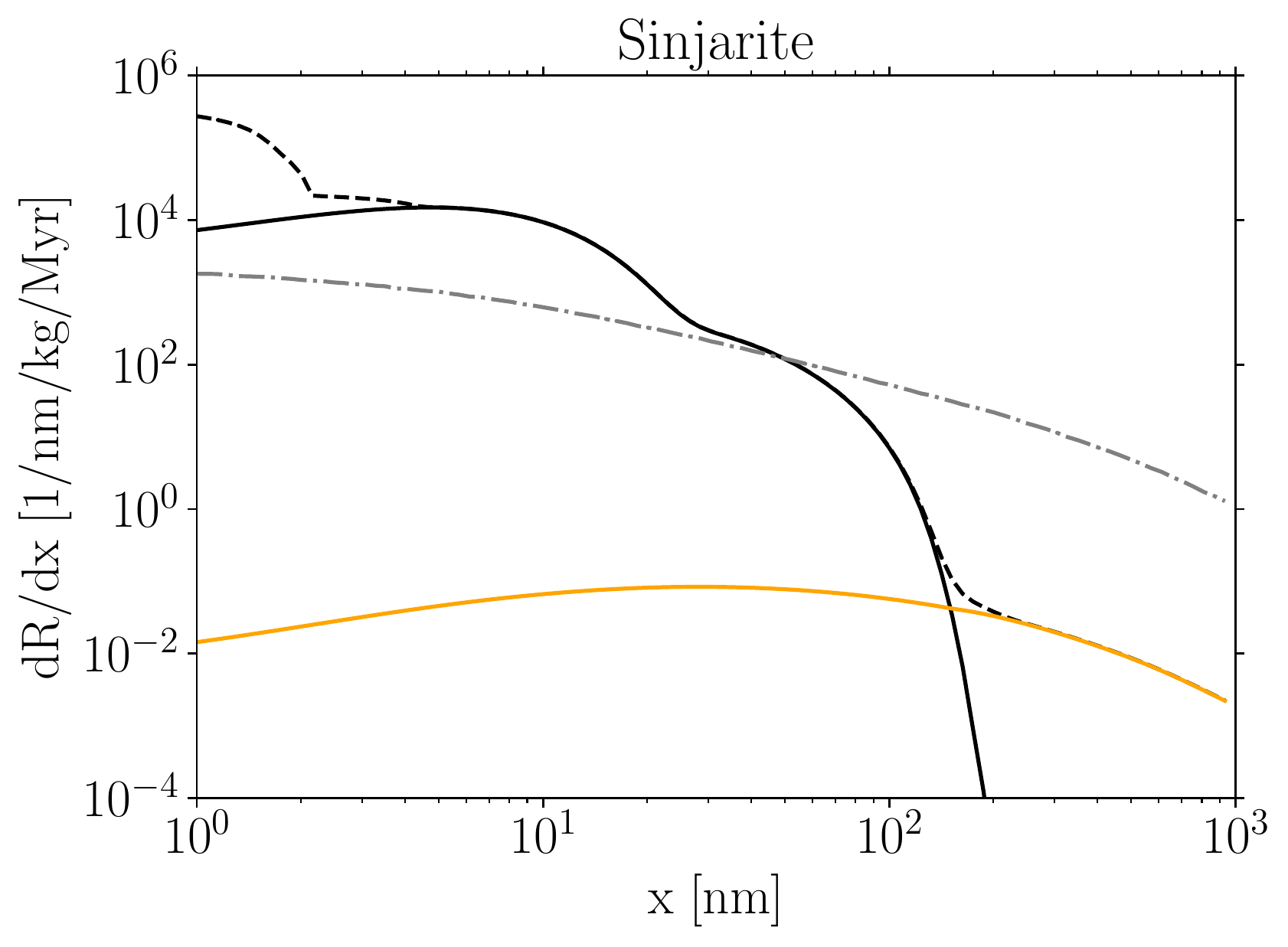}
    \includegraphics[angle=0,width=.49\textwidth]{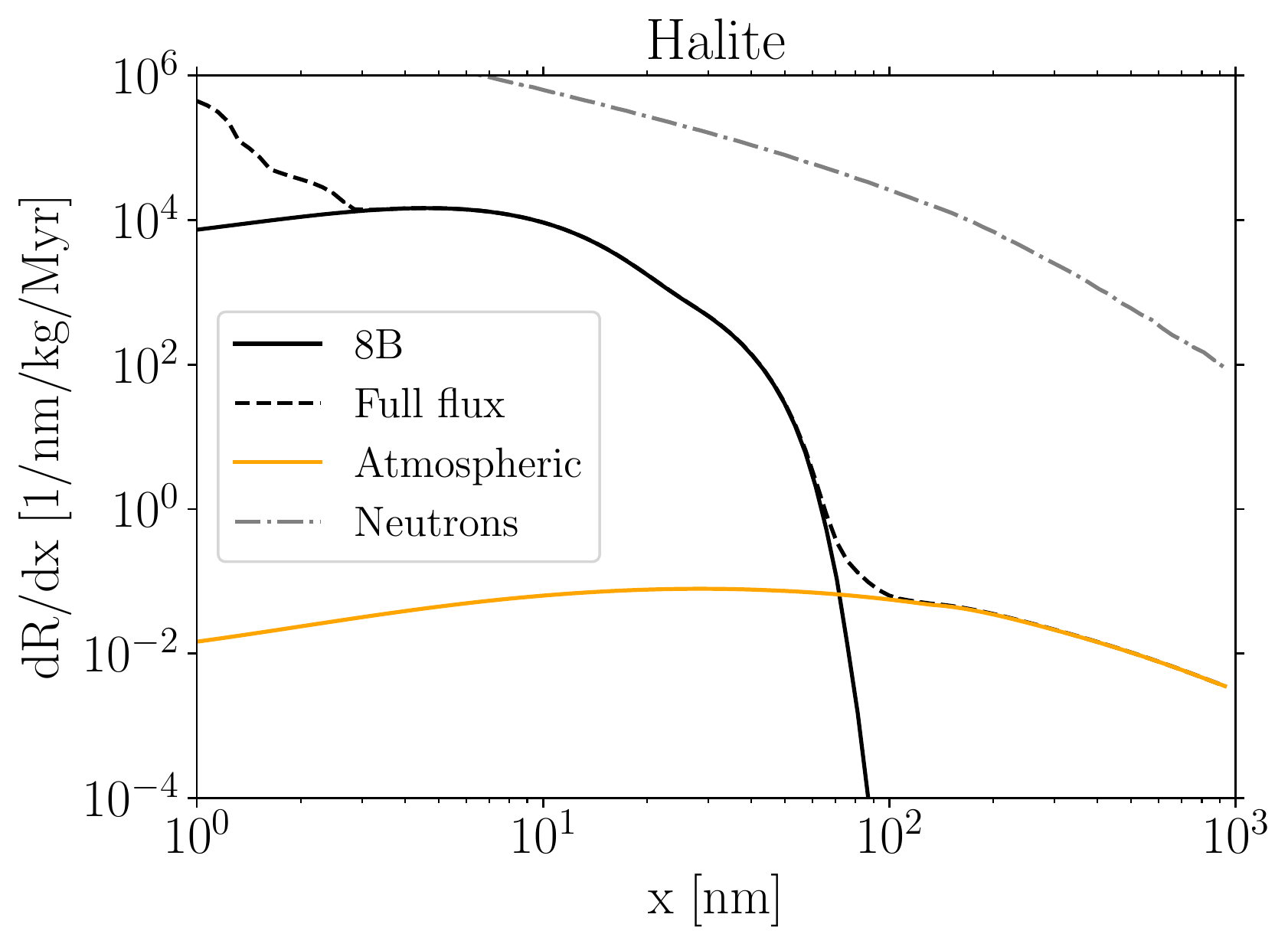}
    \caption{Track length spectra for sinjarite (left panel) and halite (right panel). Shown are the tracks induced by neutrinos and neutrons, as labeled. The tracks caused by the $^8$B flux and atmospheric neutrinos are shown separately for clarify. The central peak above tens of nm is induced by the $^8$B flux, while above around $200$ nm the tracks are induced by the atmospheric neutrino flux, while below 2--3 nm the tracks arise from the contributions of multiple solar neutrino components with lower energies than $^8$B.  
    }
    \label{fig:$^8$B_flux+bakcground_comp}
\end{figure*}

When an incoming particle collides with a target rock crystal, it is slowed down and eventually stopped, during which time its energy is deposited in the form of tracks, that is, permanent damages made in the rock. The shape and length of these tracks allow one to reconstruct the mass, energy and even direction of the incoming particle \cite{ILIC2003179}. The formation of tracks is dependent on the response of the material, which in turn depend on the chemistry of the material. 
In this section, we first discuss our material selection, and we review the computation of tracks formed by solar neutrinos and backgrounds. For more details, we refer the reader to, e.g.,  Refs.~\cite{BAUM2020135325,Edwards:2018hcf,PhysRevD.99.043541,PhysRevD.99.043014}.

In analogy to direct dark matter detection experiments, low radioactive contamination is a crucial part of material selection. 
The major consideration is the material's Uranium concentration. 
There are two processes by which ${}^{238}$U could leave tracks: 
$\alpha$ decay and spontaneous fission. 
Spontaneous fission of ${}^{238}$U nuclei gives rise to approximately two $\sim 1 \, \text{MeV}$ fast neutrons. Neutrons are also produced in ($\alpha$, n) reactions, where nuclei absorb an incident $\alpha$ particle and emit fast neutrons. 
These neutrons act similarly to solar neutrinos in their formation of tracks in the material. Cosmic rays can also cause potential background track events. However, they can be mitigated by using paleo detector material from deep underground. The primary concern will be neutrons arising from cosmic-ray muons interacting with nuclei near the target material. By a depth of $\sim 6$ km however, the estimated neutron flux is $\sim 10 \, {\rm cm^{-2} \, Gyr^{-1}}$ \cite{Mei:2005gm}. As we discuss below, we consider targets of 0.1 kg in mass corresponding to cross-sectional areas of $\sim 10 \, {\rm cm^{2}}$, implying background from cosmic-ray induced neutrons will be negligible. For a detailed description of the neutron backgrounds, see Ref.~\cite{PhysRevD.99.043014}.

We follow Ref.~\cite{PhysRevD.99.043541} and start with four minerals (olivine, nchwaningite, halite, and sinjarite), chosen because of their low levels of natural radioactive contamination. 
Olivine and Nchwaningite are examples of ultra basic rocks (UBR), which stem from the Earth's mantle \cite{Bowes1990} and are more radiopure than the average Earth crust. We follow Ref.~\cite{PhysRevD.99.043541} and adopt a Uranium concentration of $0.1$ parts per billion for UBRs. Halite and Sinjarite are marine evaporites (ME) formed after extreme evaporation of seawater \cite{Dean2013} and are even more radiopure. We follow Ref.~\cite{PhysRevD.99.043541} and adopt Uranium concentrations of $0.01$ parts per billion for MEs. 

Furthermore, the difference in the chemical composition of these minerals also has two important implications: first, they impact the rate of neutron background recorded, and second, they impact the response, i.e., the length distribution of tracks given a neutrino spectrum.
As we will show, hydrogen is a good moderator of fast neutrons, 
making minerals containing hydrogen attractive as solar neutrino detectors. 
Two of the minerals we consider contain hydrogen: nchwaningite (${\rm Mn}_{2}^{2 +} {\rm Si O}_{3} ({\rm OH})_{2} \cdot ({\rm H}_{2}{\rm O}) $), a type of UBR, and sinjarite (${\rm CaCl}_{2} \cdot 2({\rm H}_2 {\rm O})$), a type of ME. 

We base our estimates of track formation on the calculations of Ref.~\cite{Edwards:2018hcf}. 
Here, we review the main ingredients and describe how we adapt it to compute tracks from a time-dependent solar neutrino flux. 
The ionization track length for a recoiling nucleus $T$ with initial recoil energy $E_{R}$ is
\begin{equation}
    x_{T} (E_{R}) = \int_{0}^{E_{R}} dE \left( \frac{dE}{dx_{T}} (E) \right)^{-1},
\end{equation}
where $dE/dx_T$ is the stopping power for a recoiling nucleus $T$ incident on an amorphous target. If the target material has different components, the stopping power is the sum of the different contributions $V$
\begin{equation}
    \frac{dE}{dx_{T}}= \sum_V \left( \frac{dE}{dx} \right)_{V T}. 
\end{equation}
The stopping power is obtained from the SRIM code \cite{ZIEGLER20101818}, which differs at the $10\%$ level from analytical calculations \cite{BAUM2020135325}. 
Track formation from ions
with charge $Z \gtrsim 10$ and recoil energies larger than a few keV are well studied, but the situation is less clear for lighter ions~\cite{PhysRevD.99.043014}. Thus, we do not include hydrogen as a potential source of tracks, and we show in the appendix results when  hydrogen induced tracks are included.

In the search for solar neutrinos, fast neutrons pose a significant background. We model the neutron induced recoil spectra following Ref.~\cite{PhysRevD.99.043541}, which in turn is based on the {\tt sources-4A} code \cite{10.1093/rpd/nci260}, for the spontaneous fission and $(\alpha, n)$ contributions to the background. 
The code uses the parameters of 43 actinides and their half lives and spontaneous fission branches, to obtain the spontaneous fission spectra of neutrons. The spontaneous fission neutron spectra are approximated by a Watt fission spectrum. For the $(\alpha, n)$ channel, an isotropic neutron angular distribution in the centre-of-mass system is assumed. The spectra can be computed using this set up, see, e.g., Ref.~\cite{osti_15215}. The average number of neutrons must be known, which is supplied by the assumed Uranium concentration. In \Cref{fig:$^8$B_flux+bakcground_comp}, we show the neutron-induced track length spectra for sinjarite and halite with assumed Uranium concentrations of 0.01 parts per billion using dot-dashed grey lines.

Next, we compute the track length spectra induced by neutrinos. We consider the differential solar neutrino flux to study the tracks left during different time epochs of the Sun's history. The neutrino induced differential recoil spectrum per unit mass of target nuclei $T$ is given by 

\begin{equation}
    \left( \frac{d R}{d E_{R}} \right)_T = \frac{1}{m_T} 
    \int_{E_{\nu}^{min}} d E_{\nu} 
    \int_{\Delta t} d t
    \frac{d \sigma}{d E_R} \frac{d^2 \Phi_{\nu}}{d E_{\nu} dt},
    \label{eq:track_length_spectrum}
\end{equation}
where $E_{\nu}^{min} = \sqrt{m_{T} E_{R}/2}$ is the minimum energy required in order to produce a nuclear recoil with energy $E_{R}$ 
and $d^2\Phi/dEdt$ is the differential solar neutrino flux. We adopt the time-dependent $^8$B flux as described in Section \ref{sec:solar}, keeping the spectral shape of neutrinos fixed. Contrary to the flux which depends very strongly on the temperature, the spectral corrections due to thermal broadening are minimal. In practice, the code of Ref.~\cite{Edwards:2018hcf} assumes a constant solar neutrino flux, so we re-scale the track predictions by the ratio of integrated flux of neutrinos over time bins of interest. The differential cross section for coherent neutrino-nucleus scattering is

\begin{equation}
  \frac{d \sigma}{d E_{R}}  (E_{R},E_{\nu})= \frac{G_{F}^{2}}{4 \pi} \, Q_{W}^{2} m_{T} \left(1-\frac{m_{T}E_{R}}{2 E_{\nu}^{2}}\right) \, F^{2}(E_{R}) 
\end{equation}

where $G_F$ is the Fermi coupling constant, $F(E_R)$ is the nuclear form factor,  $Q_{W} \equiv \left(A_{T}-Z_{T}\right)- \left(1-4 \sin{\theta_{W}}^2 \right) Z_T$ and $\theta_{W}$ is the weak mixing angle. 
Finally, the track length spectra is obtained by summing over the target nuclei in the mineral, 
\begin{equation}
    \frac{dR}{dx}=\sum_{T} \xi_{T} \frac{d E_{R}}{dx_{T}} \left( \frac{d R}{d E_{R}} \right)_T,
\end{equation}
where $\xi_{T}$ is the mass fraction of each constituent nuclei $T$.

There are other sources of neutrinos that could leave tracks over our track lengths of interest. One example is atmospheric neutrinos. 
We compute this using the Honda atmospheric neutrino model \cite{PhysRevD.83.123001}. Another example is the diffuse supernova neutrino background (DSNB) \cite{OHare:2016pjy}, which we do not explicitly consider due to their much smaller relative contributions \cite{Edwards:2018hcf}. In fact, the neutrinos from supernovae in the Milky Way galaxy yields a higher flux over geological times than the DSNB \cite{PhysRevD.101.103017}, yet even that is lower than the $^8$B neutrino flux by orders of magnitude in the 1--10 MeV energy range. 

In \Cref{fig:$^8$B_flux+bakcground_comp}, we show the neutrino-induced track spectra, highlighting the different neutrino components, for sinjarite (left panel) and halite (right panel). The longest tracks, left by the highest energy neutrinos, are unsurprisingly induced by atmospheric neutrinos, shown by the solid orange curves. For halite, the neutrino induced tracks are however overwhelmed by the much more numerous neutron induced tracks. 
On the other hand, for sinjarite the neutrino induced tracks exceed the neutron background tracks below track lengths of $\sim 30$ nm. 
Below 10 nm, the tracks are dominated by a mixture of multiple solar neutrino components, including $^7$Be, pep, and CNO. 
Therefore, at intermediate track lengths between 10--30 nm, there is a window where the $^8$B neutrino flux is the dominant source, exceeding also the neutron-induced background. 

\begin{figure*}[t!]
    \centering
    \includegraphics[angle=0,width=.49\textwidth]{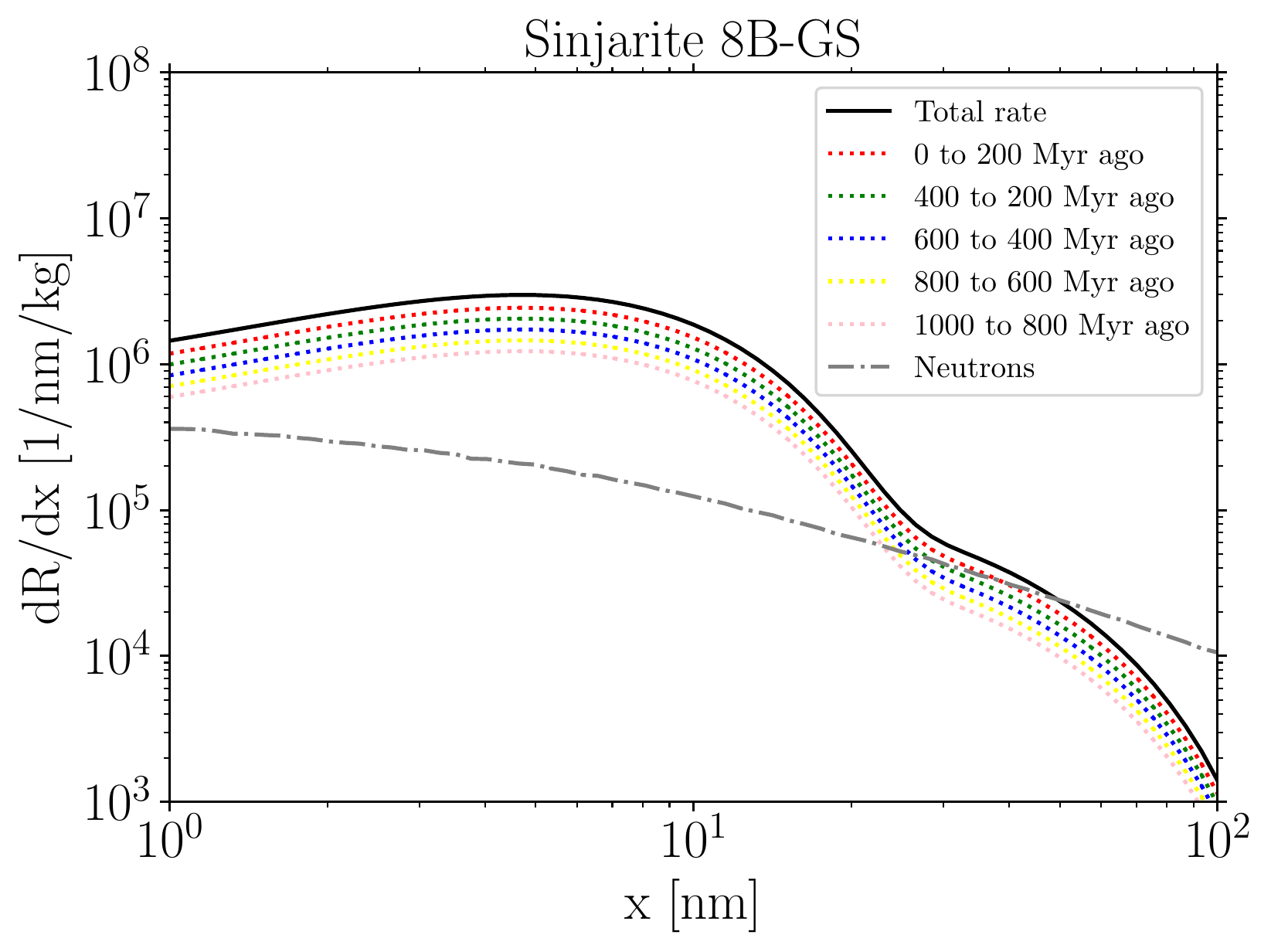}
    \includegraphics[angle=0,width=.49\textwidth]{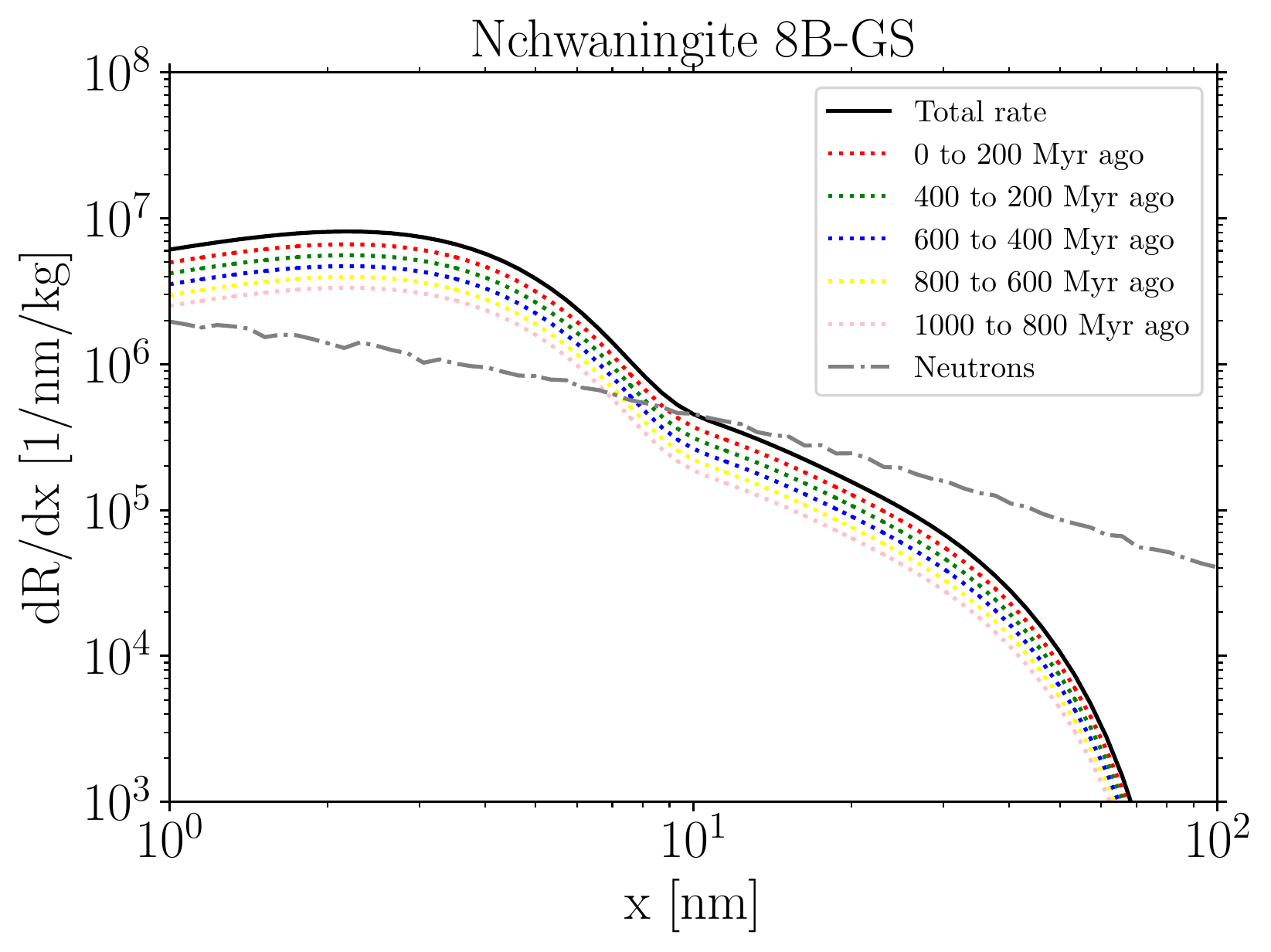}
    \caption{Same as Fig.~\ref{fig:$^8$B_flux+bakcground_comp} but showing the tracks induced by the $^8$B flux over time bins of 200 Myr widths. The neutron background from radioactive processes can be seen in gray dashed line and remains the same for every 200 Myr bin. 
    }
    \label{fig:rate_8B_tracks_myr_GS_2_Nch}
\end{figure*}

The tracks spectrum induced by solar neutrinos in sinjarite show a kink at around 30 nm. This is not due to a feature in the neutrino spectrum, but rather, is associated with the transition in tracks left by different nuclei of the sinjarite crystal (${\rm CaCl}_{2} \cdot 2({\rm H}_2 {\rm O})$). 
Below track lengths of $\sim 30 $ nm, the contribution is dominated by recoils of Cl and Ca, whereas above $\sim 30 $ nm, O recoils dominate. This is not because of the number of tracks it would create, but because of their recoil kinematics and the resulting tracks lengths they reach.

The kink is not as apparent in halite (see right panel of \Cref{fig:$^8$B_flux+bakcground_comp}) and olivine which have a different chemical composition and the transition between the different nuclei of each mineral component is much softer.

In \Cref{fig:rate_8B_tracks_myr_GS_2_Nch}, we show the track spectra in 200 Myr time bins, for sinjarite (left panel) and nchwaningite (right panel). These are two materials with a large detection window for $^8$B neutrino induced tracks. The composition of these minerals is what makes them good candidates for our study, in contrast to Olivine and Halite; see also Fig.~1 of Ref.~\cite{Edwards:2018hcf}.

\section{Results}

\subsection{Time variation of number of events}

We start our analysis with the detectability of the time-evolution of tracks induced by the time-dependent solar neutrino flux. This requires measuring the total number of tracks in multiple mineral samples of different ages. We consider paleo detectors with ages up to 1 Gyr with 200 Myr age accuracy. It is expected tracks would be preserved for around 1 Gyr, but beyond this time scale, they could fade due to annealing of the rocks under the high temperatures encountered deep underground \cite{PhysRevD.99.043014}. 
Our choice of 200 Myr time bins is motivated by the precision of techniques to date rock samples. The most relevant method for dating Gyr-aged rocks is fission track dating, which is expected to be reliable at the level of $\sim 10$\% \cite{doi:10.1146/annurev.earth.26.1.519}.

\begin{figure*}[t]
\centering
\includegraphics[angle=0,width=.49\textwidth]{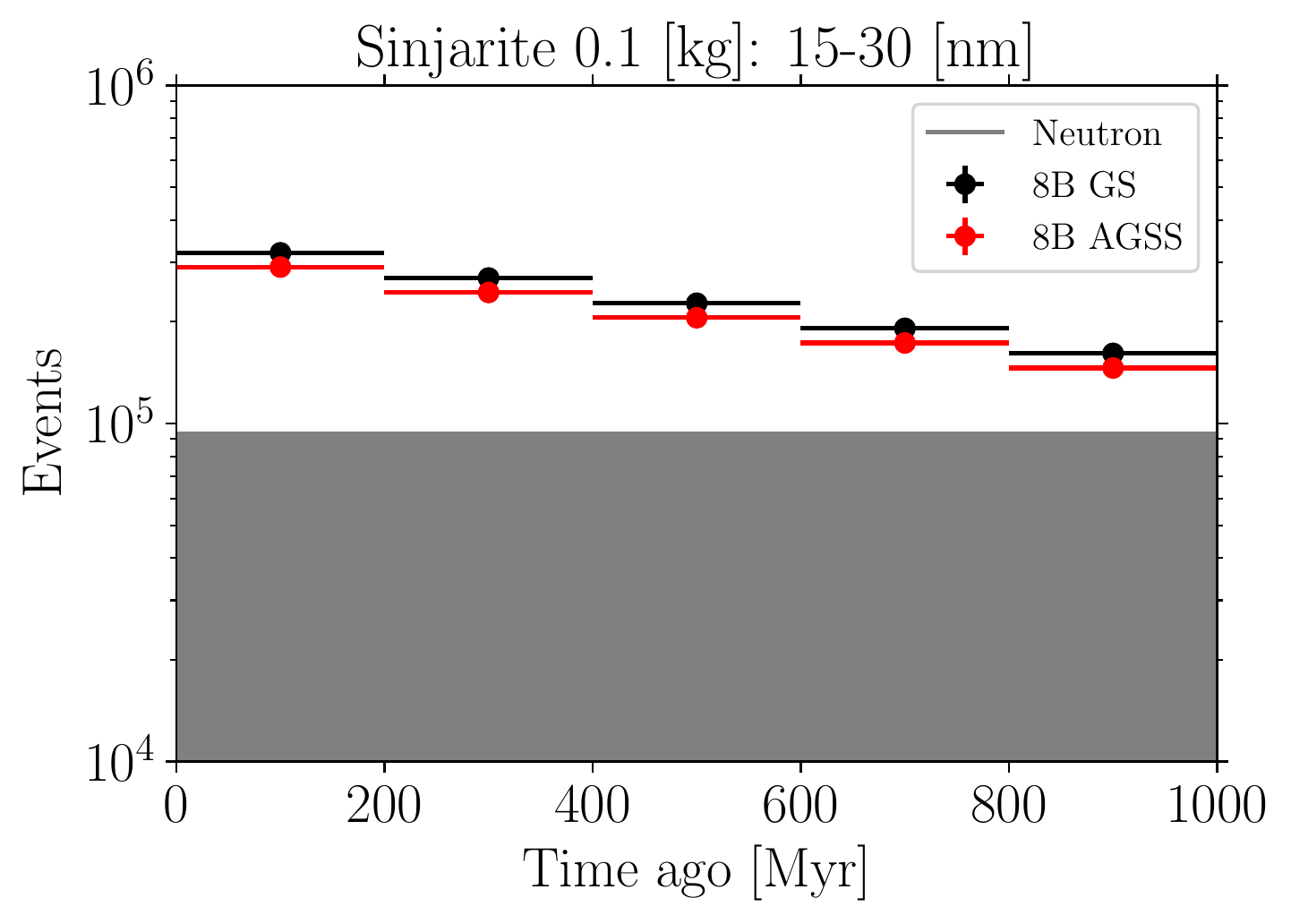}
\includegraphics[angle=0,width=.49\textwidth]{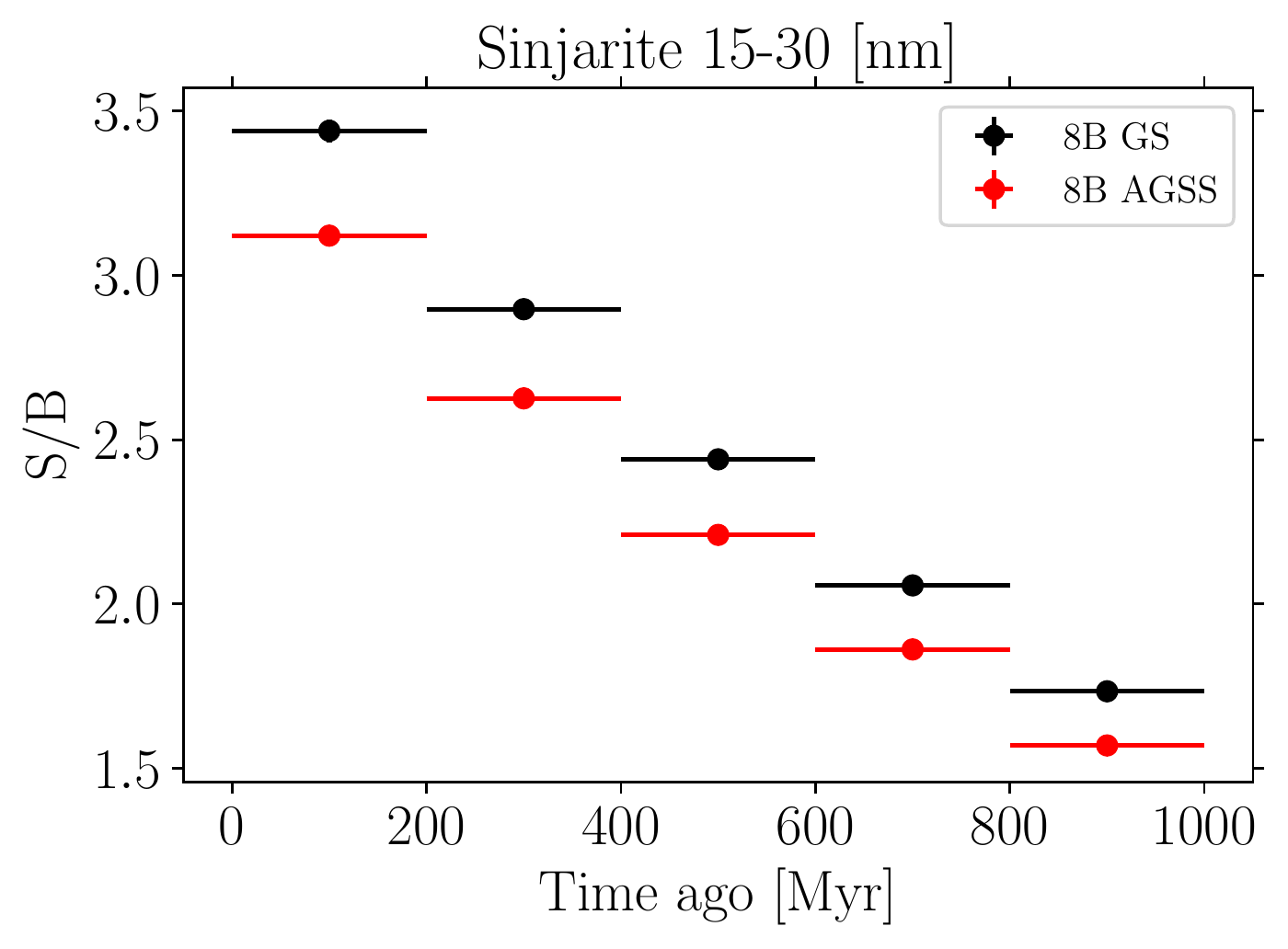}
\caption{Left: Number of events per 200 Myr time bins, for 0.1 Kg of Sinjarite. Events are summed over track lengths of 15 to 30 nm. The black dots represent the GS (reference) SSM and red represents AGSS SSM. The shaded region is where the neutron backgrounds will dominate the events. Right: signal-to-background ratio, separately for the GS and AGSS models. The time binning is the same as on the left panel.}
\label{fig:Sinjarite_15_30nm_200my_01kg}
\end{figure*}

We consider different exposure scenarios following Ref.~\cite{PhysRevD.99.043541}. 
Our default assumption is a mineral of mass of 0.1 kg and a minimal track length resolution of 15 nm. This corresponds to the assumed resolution and mass of the ``high exposure'' case of Ref.~\cite{PhysRevD.99.043541}. Note that this implies exposures of $\varepsilon \sim 20\,\text{Kg Myr}$ for each of our 200 Myr time bins. 

The tracks which have persisted over time could be read out with helium ion beam microscope, with spatial resolution of $\sim $ nm and able to image up to depths of $O(100)$ nm \cite{PhysRevD.99.043014}. 
Small angle X-ray scattering can achieve $\sim 15 $ nm three dimensional spatial resolution, which is our adopted minimum length sensitivity, and it can measure up to $O(100)$ g of material, also our choice of sample mass. 
The peak of the $^8$B track length spectra actually lies at lower lengths ($ \sim 7$ nm for sinjarite), which would provide a more favorable signal over background ratio than our 15 nm minimum track length. However, measuring tracks as small as $7 \, \text{nm}$ will need technology development and we conservatively assume the more established 15 nm resolution. In the Appendix, we show results for a 10 nm resolution case.

\begin{figure*}[hbt!]
\centering
\includegraphics[angle=0,width=.49\textwidth]{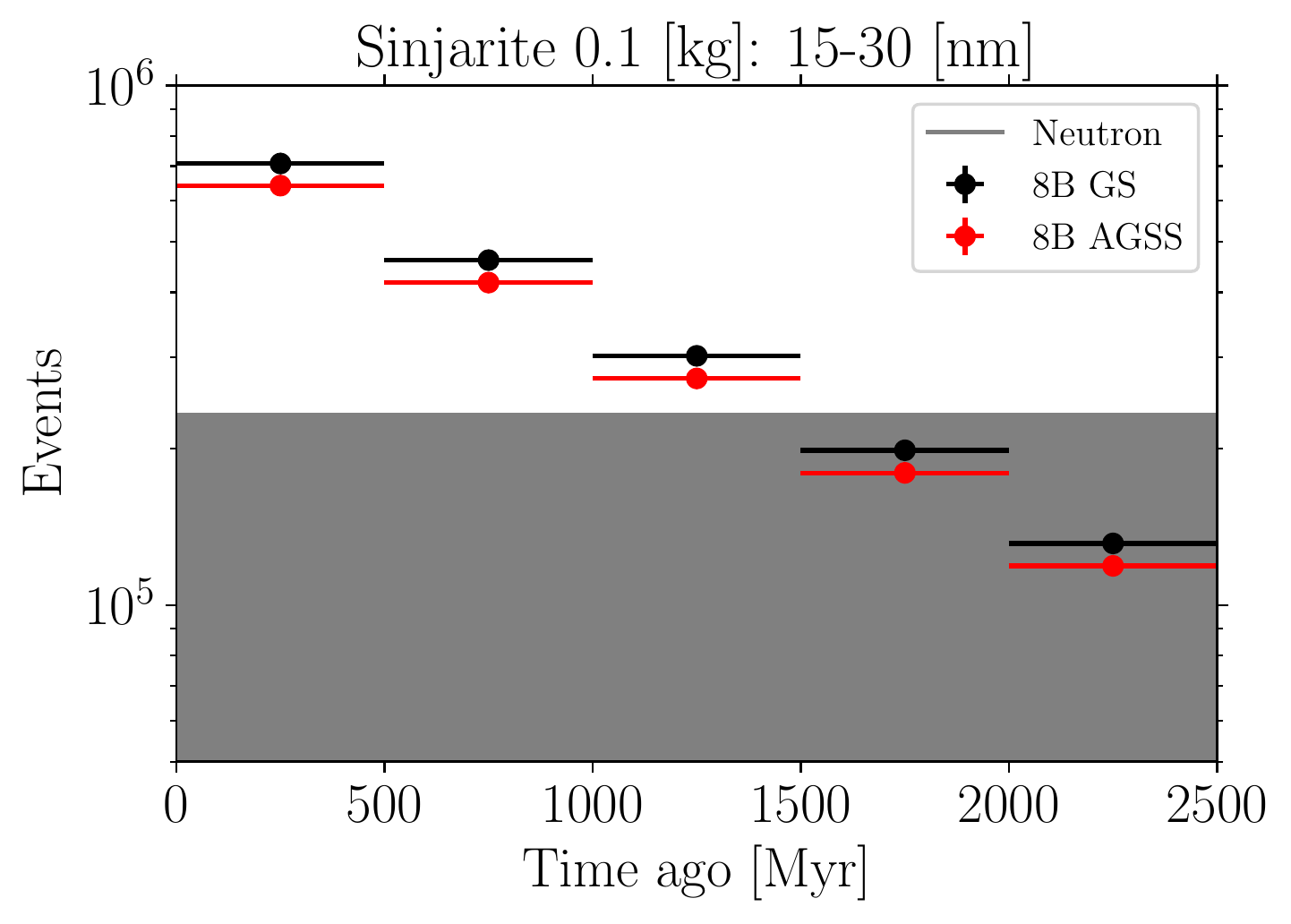}
\includegraphics[angle=0,width=.49\textwidth]{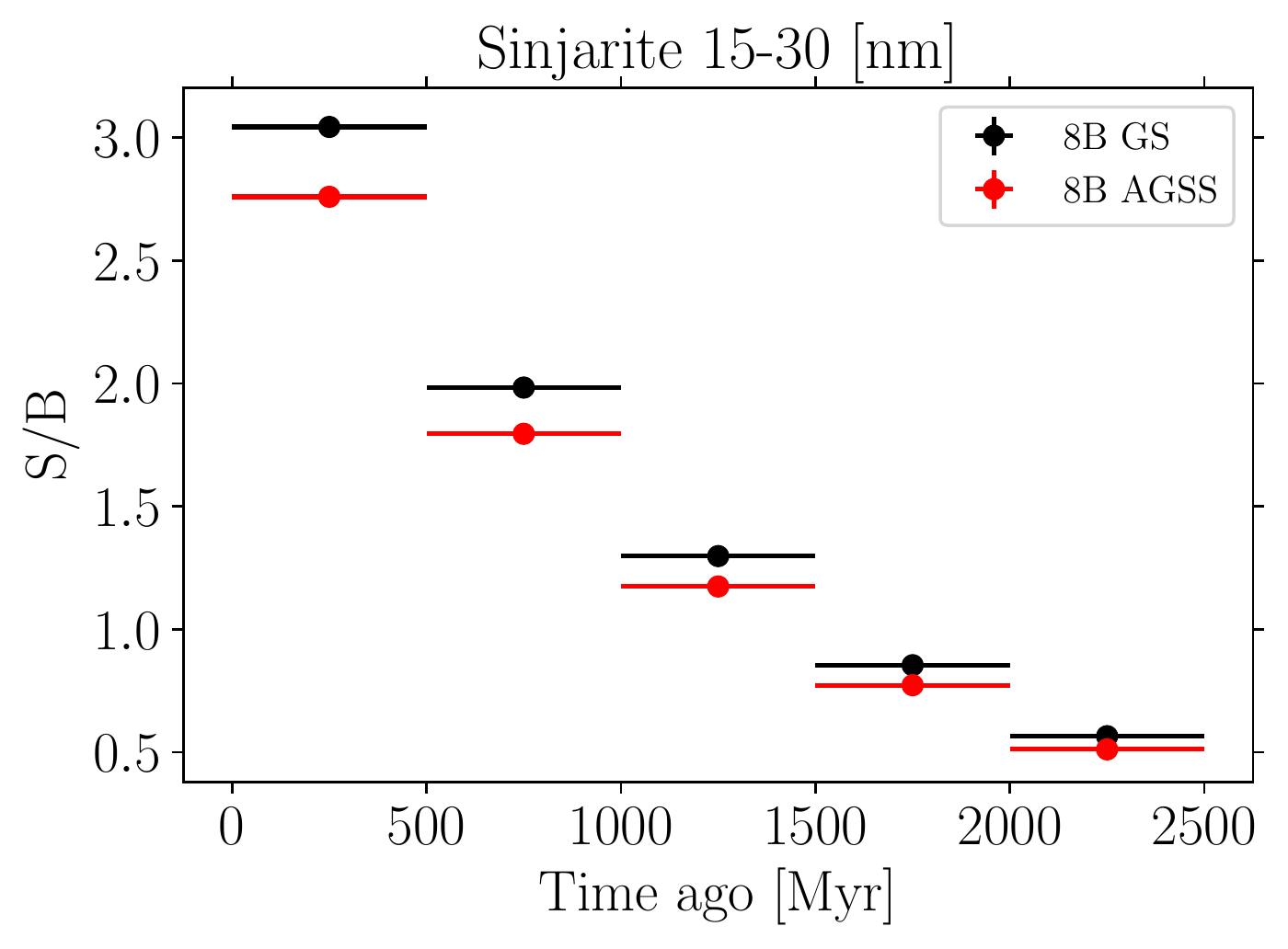}
\caption{The same as Fig.~\ref{fig:Sinjarite_15_30nm_200my_01kg}, but extending out to 2.5 Gyr and with wider 500 Myr time bins.}
\label{fig:Sinjarite_15_30nm_500my_01kg.pdf}
\end{figure*}

To maximize the signal to noise, we consider a maximum track length of 30 nm. As can be seen in the left panel of  \Cref{fig:rate_8B_tracks_myr_GS_2_Nch}, this helps to mitigate the neutron background which starts exceeding the $^8$B neutrino events above 30 nm. In the case of nchwaningite (\Cref{fig:rate_8B_tracks_myr_GS_2_Nch} right panel), the peak is more pronounced around track length of 2 nm, and by 15 nm the ratio of signal over background is already disfavoured. 
This difference is caused by the the different nuclear compositions which in turn determine the recoil response to the same $^8$B neutrino spectrum.
Thus, we find a strong preference for sinjarite over nchwaningite, 
even though they both share the benefit of containing hydrogen in their molecular structures.
We note that in these estimates we assume hydrogen do not efficiently leave tracks. In the appendix, we discuss how the results numerically change if they are included.

Our main results, for two metallicity models and different time windows, are shown in \Cref{fig:Sinjarite_15_30nm_200my_01kg} where we show the total number of events per time bin (left panel) and the computed signal-to-noise per time bin (right panel), both for sinjarite. 
We see that in every time bin, the signal over background ratio is above 1, reaching $\sim 3.5$ for the most recent past. 

Going beyond $\sim 1$ Gyr in the past poses additional difficulties. In addition to the effects of heat annealing, the signal become overwhelmed by backgrounds. In \Cref{fig:Sinjarite_15_30nm_500my_01kg.pdf} we show the
reach up to 2.5 Gyr in 500 Myr time bins. 
We see that beyond 1--1.5 Gyr the solar neutrino induced tracks fall below the neutrino signal, and measuring the solar neutrino tracks become difficult. 
However, the situation improves if a higher resolution can be achieved. 
In the Appendix we consider the case of a track resolution of 10 nm. 
An improvement with respect to our 15--30 nm case is expected from \Cref{fig:rate_8B_tracks_myr_GS_2_Nch}, but it has a qualitatively important impact for measuring the solar activity beyond the 1 Gyr age, provided paleo detectors of those ages can be found.

\subsection{Sensitivity to metallicity model}

For our 200 Myr time bin width result (\Cref{fig:Sinjarite_15_30nm_200my_01kg}), the difference between the GS and AGSS metallicity models is $\sim 10^{4}$ tracks per time bin, i.e., approximately $\sim 10$\% of the track counts. As can be seen in the panels, the difference between GS and AGSS is more or less time independent. Thus, we can sum over the entire 1 Gyr range for maximum statistics. The total number of tracks is $\sim 1.17 \times 10^{6}$ and $\sim 1.06 \times 10^{6}$, for GS and AGSS, respectively. If we assume only Poisson errors, then $\sigma \sim 10^3$, and a difference of 10\% due to the metallicity models would be easy to measure.

In reality, the measurement will be dominated by systematic effects. So far, we have assumed the major source of background arising from neutron tracks is perfectly known. The uncertainty of this background is related to the initial concentration of radioactive material present in the crystal. The initial radioactivity can then be estimated by the number of whole decays chains identified through long tracks. Thus, the normalization of the neutron background is likely to be estimated to high precision.

Nevertheless, to study the potential impacts of any uncertainty on the mineral's  radioactivity, we assume two situations: a $1\%$ systematic uncertainty in the neutron background, as well as up to $10\%$ uncertainty. 
In both cases, it would be possible to distinguish the two metallicity models. For example, for the 10\% case, the total number of tracks over the 1 Gyr window in the GS and AGSS metallicity models is $(1.63 \pm  0.05)\times10^{6}$ and $(1.52 \pm 0.05)\times 10^{6}$  tracks, respectively. Here, the numbers include the number of tracks caused by fast neutron ($\sim 5 \times 10^5$) and the error bars represent a $\pm 10$\% uncertainty of the neutron induced tracks. Even with this generous uncertainty, it would be possible to gain insight into the metallicity models based on total number of tracks.

\section{Conclusions}

We have considered SSMs with two metallicities (GS and AGSS) to study the detectability of the boron-8 ($^8$B) solar neutrino flux over gigayear timescales using paleo detectors. 
Our default setup is a paleo detector composed of a collection of 0.1 kg of sinjarite crystals of different ages, up to 1 Gyr old and aged with 200 Myr resolution, but we also consider an older sample reaching 2.5 Gyr with 500 Myr age-dating resolution. We identified track lengths of 15--30 nm to be the most suitable for detecting the $^8$B neutrino flux with established technologies. We found that up to 1--1.5 Gyr, the $^8$B signal to background ratio is favourable for measuring the time-evolution of the $^8$B neutrino flux and for distinguishing the two metallicity models. 

The main background to solar neutrinos is caused by fast neutrons, originating from radioactive sources within the minerals. 
The normalization of this background solely depends on the original concentration of radioactive materials. We follow Ref.~\cite{PhysRevD.99.043014} and adopt a Uranium concentration of 0.01 parts per billion for sinjarite. 
Since the neutron background rate scales with this concentration, the detectability of the $^8$B neutrino flux with sinjarite requires the concentration to be not more than a few times 0.01 parts per billion.
In principle this concentration can be measured, e.g., using the number of full $^{238}$U decay chains in the target mineral as probed by longer tracks, and therefore constrained accurately. We follow Ref.~\cite{PhysRevD.99.043014} and assume a normalization uncertainty of 1\%. 
Nevertheless, we also check up to $10\%$ uncertainty and we find that the two metallicity solar models can be differentiated even with the larger background uncertainty. 

Among the paleo detector materials discussed in the literature (e.g., \cite{PhysRevD.99.043541,Edwards:2018hcf}), we identify sinjarite as uniquely optimal for solar neutrino studies. Its hydrogen component helps to reduced neutron backgrounds by slowing them and reducing the numbers of tracks related to this source. Nchwaningite is another material studied in the literature \cite{Edwards:2018hcf} and like sinjarite contains a hydrogen component. It is a good candidate with good signal to background ratios for shorter tracks. This mineral would be competitive in the future when technology and theory will allow measurements and interpretations of $<10$ nm tracks. 

To conclude, paleo detectors represent a novel and intriguing opportunity to uniquely probe solar neutrinos on gigayear timescales. This could open a new window to inform the Solar Standard Model, as well as measure the past solar activity and guide inputs for solar system planetary climates.

\acknowledgments

We thank Sebastian Baum and Patrick Huber for fruitful discussions and to the authors of Ref.~\cite{Edwards:2018hcf} for making their code publicly available. The work of N.T-A\ is supported by the U.S.\ Department of Energy under the award numbers DE-SC0020250 and DE-SC0020262. The work of S.H.\ is supported by the U.S.\ Department of Energy Office of Science under award number DE-SC0020262 and NSF Grants numbers AST-1908960 and PHY-1914409.

\appendix*
\section{Other Results}

\begin{figure}[t!]
\centering
\includegraphics[angle=0,width=.49\textwidth]{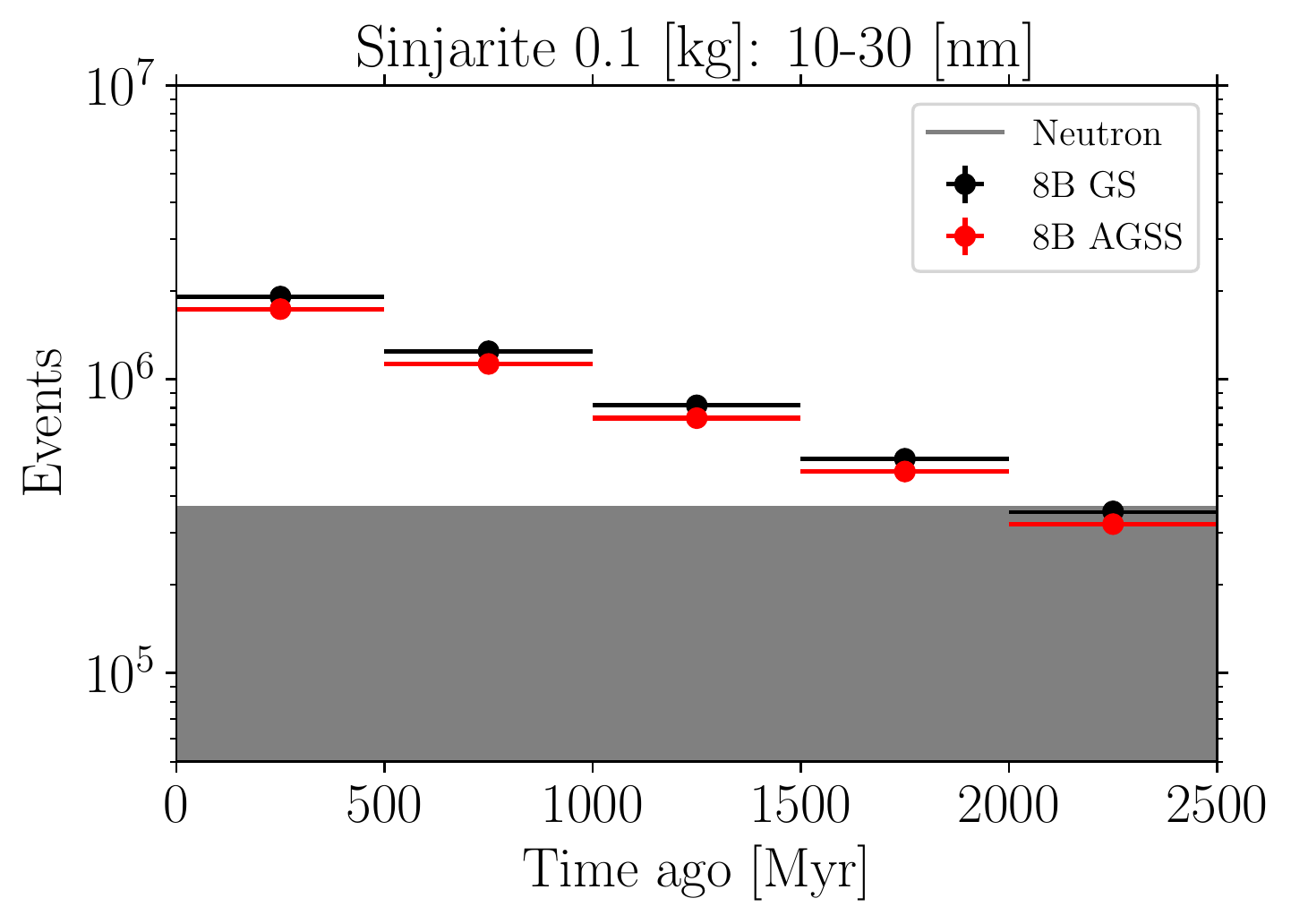}
\caption{Same as Fig.~\ref{fig:Sinjarite_15_30nm_500my_01kg.pdf}, but summed over a wider track lengths of 10--30 nm.}
\label{fig:Sinjarite_10_30nm_200my_SB_error}
\end{figure}

Here, we explore additional cases. First, we consider a track length sensitivity as short as 10 nm. This is a more optimistic scenario compared to our default 15 nm, and will require R\&D for better measuring instruments. 
Exploring shorter track lengths implies, in terms of WIMP Dark Matter masses for example, going to smaller masses. Similarly, for solar neutrino search it opens new track length regimes to perform higher signal to background searches. We show the results in \Cref{fig:Sinjarite_10_30nm_200my_SB_error}, which is to be contrasted with \Cref{fig:Sinjarite_15_30nm_500my_01kg.pdf}. 
Here, we also take 500 Myr time bins and 0.1 Kg of material. The gains due to the improved sensitivity become noticeable over these few Gyr timescales. For example, up to 2 Gyr the signal over background ratio is favourable, which was not possible when the track length sensitivity is 15 nm. 

Next, we also check the time evolution of the same mineral, sinjarite, but now considering also tracks formed by hydrogen recoils. This channel is neglected in our default predictions due to the uncertain nature of tracks left by low-$Z$ ions. The hydrogen recoils increase both the signal and background tracks, but in our search window the increase in the neutron background is more important. Due to kinematics, the signal increase occurs at higher track lengths (over $\sim 200$ nm in the case of sinjarite). Thus, adding its contributions does not make a big difference in the signal prediction between 15--30 nm. On the other hand, the neutron background sees an increase over a much wider range of track lengths, including our previously identified signal window 15--30 nm. As a result, the background starts being competitive to the signal above track lengths  as small as $\sim 20$ nm.

\begin{figure}[hbt!]
\centering
\includegraphics[angle=0,width=.49\textwidth]{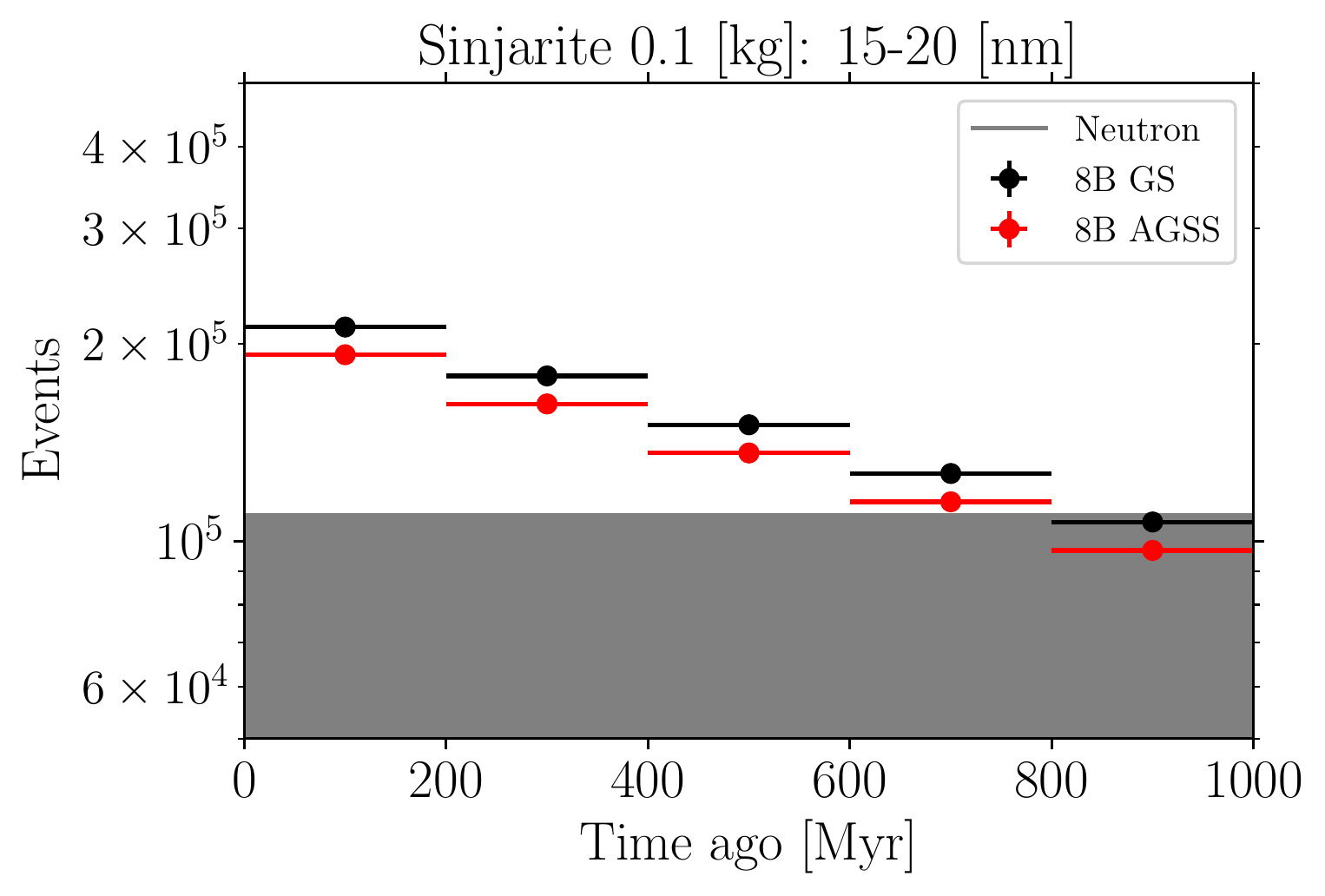}
\caption{Same as Fig.~\ref{fig:Sinjarite_15_30nm_200my_01kg}, but including hydrogen induced tracks and summed over a narrower track lengths of 15--20 nm.}
\label{fig:Sinjarite_15_20nm_200my_01kg}
\end{figure}

We show the time evolution for sinjarite including its hydrogen induced track contribution, using only 15--20 nm in Fig.~\ref{fig:Sinjarite_15_20nm_200my_01kg}. We can see that compared to Fig.~\ref{fig:Sinjarite_15_30nm_200my_01kg}, the signal becomes smaller than the background as early as $\sim 800$ Myr. If instead we continue to use 15--30 nm, the signal becomes smaller than the background already by $\sim 200$ Myr. 

\bibliography{references}

\end{document}